\documentclass[journal]{IEEEtran}
\usepackage{amsmath}
\usepackage[euler]{textgreek}
\usepackage{adjustbox,lipsum}
\usepackage{graphicx}
\usepackage[justification=centering]{caption}
\usepackage[section]{placeins}
\usepackage{float}
\usepackage[backend=biber,style=authoryear]{biblatex}
\usepackage{comment}

\begin{document}
%
\title{A Study of Revenue Cost Dynamics in Large Data Centers: A Factorial Design Approach}
%
%
%

\author{Gambhire~Swati~Sampatrao-\IEEEmembership{Member~IEEE,}~ Sudeepa~Roy~Dey-\IEEEmembership{Member~IEEE,}~Bidisha~Goswami-\IEEEmembership{~Member~IEEE}~Sai~Prasanna~M~S-\IEEEmembership{Member~IEEE,} ~Snehanshu~Saha-\IEEEmembership{Senior~Member~IEEE} }  

\maketitle

\begin{abstract}
Revenue optimization of large data centers is an open and challenging problem. The intricacy of the problem is due to the presence of too many parameters posing as costs or investment. This paper proposes a model to optimize the revenue in cloud data center and analyzes the model, revenue and diiferent investment or cost commitments of organizations investing in data centers. The model uses the Cobb-Douglas production function to quantify the boundaries and the most significant factors to generate the revenue. The dynamics between revenue and cost is explored by designing an experiment (DoE) which is an interpretation of revenue as function of cost/investment as factors with different levels/fluctuations. Optimal elasticities associated with these factors of the model for maximum revenue are computed and verified . The model response is interpreted in light of the business scenario of data centers.  
\end{abstract}


\IEEEpeerreviewmaketitle

\section{Introduction}
\IEEEPARstart{T}{he}
data center has a very important role in cloud computing domain. The costs associated with the traditional data centers include maintenance of mixed hardware pools to support thousand of applications, multiple management tools for operations, continuous power supply, facility of water for cooling the power system and network connectivity, etc. These data centers are currently used by internet service providers for providing service such as infrastructure and software. Along with the existing pricing, a new set of challenges are due to the up-gradation , augmenting  different dimensions of the cost optimization problem. 

Most of the applications in the industry are shifting towards cloud system, supported by different cloud data centers. I \& T Business industries assume the data center to function as a factory-like utility that collects and processes information from an operational standpoint. They value the data that is available in real time to help them update and shape their decisions. These industries do expect that, the data center needs to be fast enough to adapt to new, rapidly deployed, public facing and internal user applications for seamless service. The technology standpoint demands the current data centers to support mobility, provisioning on demand, scalability, virtualization and the flexibility to respond to fast-changing operational situations. Nevertheless, from an economic viewpoint, a few years of edgy fiscal conditions have imposed tight budgets on IT organizations in both public and private sectors. This compels them to rethink about remodeling and smart resource management. The expectation in price and performance from clients needs to be maximum while expectation ( within the organisation)in terms of cost has to have a minimum. This is the classic revenue cost paradox. Organizations expect maximum output for every dollar invested in IT. They also face pressure to reduce power usage as a component of overall organizational strategies for reducing their carbon footprint.
Amazon Web Services(AWS) and other data center providers are constantly improving the technology and define the cost of servers as the principle component in the revenue model. For example, AWS spends approximately 57\% of their budget towards servers and constantly improvise in the procurement pattern of three major types of servers \cite{[4]} .
The challenge that a data center faces is the lack of access of basic data critical towards planning and ensuring the optimum investment in power and cooling system. Inefficient power usage, including the sub-optimal use of power infrastructure and over investment in racks and power capacity, burdens the revenue outlay of organizations. Other problems include Power and Cooling excursions i.e.  availability of power supply during pick business hours and identifying the hot-spots to mitigate and optimize workload placement based on power and cooling availability.
Since, energy consumption of cloud data centers is a key concern for the owners, energy costs (fuel) continue to rise and CO2 emissions related to this consumption have become relevant \cite{[9]} . Therefore, saving money in the energy budget of a cloud data center, without sacrificing Service Level Agreements (SLA) is an excellent incentive for cloud data center owners, and would at the same time be a great success for environmental sustainability. The ICT resources, servers, storage devices and network equipment consume maximum power.

In this paper, a revenue model with the cost function based on the sample data is proposed. This uses Cobb-Douglas production function to generate a revenue and profit model and defines the response variable as production or profit, a variable that needs to be optimized.The response variable is the output of several cost factors. The contributing factors are Server type and power and cooling costs. The proposed model heuristically identifies the elasticity ranges of these factors and uses a fitting curve for empirical verification. However, the cornerstone of the proposed model is the interaction and dependency between the response variable, Revenue or Profit against the two different types of cost as input/predictor variables.

The remainder of the paper is organized as follows. Section \ref{sec:relatedwork} discusses the related work, highlighting and summarizing diffeent solution approaches to the cost optimization problem in data center. In Section III, Revenue Optimization in Data Center is discussed. This section explains the Cobb-Douglas production function which is the backbone of the proposed revenue model. Section IV talks about DoE that used to build and analyze the model . Section V elucidates the critical factors of revenue maximization in the Cobb-Douglas model. In section VI, the impact of the identified factors in the proposed design is discussed. The detailed experimental observation on IRS,CRS and DRS is provided in Section VII . Section VIII describes  various experiments conducted for validation. Section IX is about  predictive analysis to forecast the revenue from the observation. Section X concludes our work.

\section{Related Work} \label{sec:relatedwork}
Cloud data center optimization is an open problem which has been discussed by many researchers. The major cloud providers such as Amazon, Google and Microsoft  spend millions for servers, substation power transformers and cooling equipments. Google \cite{[1]} has reported \$ 1.9 billion in spending on data centers in the year of 2006 and \$2.4 billion in 2007. Apple has spent \$45 million in 2006 for data center construction cost \cite{[2]}. Facebook spent \$606 million on servers, storage and network gear and data centers \cite{[3]}. Budget constraints force the industries to explore different strategies that ensures optimal revenue.  A variety of solutions have been proposed in the literature aimed towards reducing the cost of the data centers. Ghamkhari and Mohsenian-Rad \cite{[5]} highlight the trade-off between minimizing data center's energy expenditure and maximizing their revenue for offered services. The paper significantly identifies both the factors i.e minimization of energy expense and maximization of revenue cost. Their experimental design, however, could not present any analysis regarding  contribution of factors to revenue generation.
Chen {\em et al.} \cite{[6]} propose a model that optimizes the revenue i.e expected electricity payment minus the revenue from participatory day-ahead data response. The author proposes a stochastic optimization model identifying the constraints of other cost factors associated with data centers. This may always not be applicable to real cloud scenario where on-demand, fast response is a need and the elasticity of cost factors  has  significant contribution.
Toosi {\em et al.} \cite{[7]} have addressed the issue of revenue maximization by combining three separate pricing models in cloud infrastructure. Every cloud provider has a limitation of its resources. The authors propose a framework to maximize the revenue through an optimal allocation which satisfy dynamic and stochastic need to customersby exploiting stochastic dynamic programming model.
\cite{[8]} argues that a fine-grained dynamic resource allocation of VM in a data center improves better utilization of resources and indirectly maximize the revenue. The authors have used trace driven simulation and shows overall 30\% revenue increment. 
Another possible solution involves migration and replacement of VM's ; Zhao et al. \cite{[8]} proposed an online VM placement algorithm for enhancing revenue of the data center. The proposed framework has not discussed the power consumption of VM for communication and migration which actually has a huge impact on price. Saha {\em et al.} \cite{[10]} have proposed an integrated approach for revenue optimization. The model is utilized to maximize the revenue of service provider without violating the pre-defined QoS requirements, while minimizing cloud resource cost. The formulation uses the Cobb-Douglas production function \cite{[11]} , a well known production function widely used in economics. Available scholarly document in the public domain emphasize the need for a dedicated deployment model which meets the cost demand while maintaining profitability. 

\section{Revenue Optimization and Data Centers} \label{sec:revenue}
The Cobb-Douglas production function is a particular form of the production function[11]. The most attractive features of Cobb-Douglas are: Positively decreasing marginal product,Constant output elasticity, equal to \(\beta\) and \(\alpha\) for L and K, Constant returns to scale equal to \(\alpha\)+\(\beta\).
 \begin{equation} Q(L,K)=AL^\beta K^\alpha  \end{equation}
The above equation represents revenue as a function of two variables or costs and could be scaled up to accomodate a finite number of parameters related to investment/cost as evident from equation $(2)$.
The response variable is the outcome. e.g. Revenue output due to factors such as cost, man-hours and the levels of those factors. The primary and secondary factors as well as replication patterns need to be ascertained such that the impact of variation among the entities is minimized. Interaction among the factors need not be ignored. A full factorial design with the number of experiments equal to $$\sum_{i=1}^k n_i $$ would capture all interactions and explain variations due to technological progress, the authors believe. This will be illustrated in the section titled \textbf{Factor analysis and impact on proposed design}.
\subsection *{\textbf{A. Production Maximization}}
Consider an enterprise that has to choose its consumption bundle (S, I, P, N) where S, I, P and N are number of servers, investment in infrastructure, cost of power and networking cost and cooling respectively of a cloud data center. The enterprise wants to maximize its production, subject to the constraint that the total cost of the bundle does not exceed a particular amount. The company has to keep the budget constraints in mind and restrict total spending within this amount.\\
The production maximization is achieved using Lagrangian Multiplier. The Cobb-Douglas function is:\begin{equation} f(S,I,N,P)=kS^\alpha I^\beta P^\gamma N^\delta  \end{equation}
Let \textit{m} be the cost of the inputs that should not be exceeded.
\begin{equation}w_1S+w_2I+w_3P+w_4N=m\nonumber \end{equation}
\(w_1\): Unit cost of servers\\\(w_2\): Unit cost of infrastructure\\\(w_3\): Unit cost of power\\\(w_4\): Unit cost of network\\ \\
Optimization problem for production maximization is:

\begin{equation}max\: \:  f(S,I,P,N) \:subject\: \: to\:\: m \nonumber \end{equation}

The following  values of S, I, P and N thus obtained are the values for which the data center achieves maximum production under total investment/cost constraints. 
\begin{align}& S=\frac{m\alpha}{w_1}(1+\beta+\gamma+\delta)\\
&I=\frac{m\beta}{w_2}(1+\alpha+\gamma+\delta)\\
 &P=\frac{m\gamma}{w_3}(1+\alpha+\beta+\delta)\\
&N=\frac{m\delta}{w_4}(1+\alpha+\beta+\gamma)\end{align}
The above results are proved in Appendix 1.\\

\subsection*{\textbf{B. Profit Maximization}}
Consider an enterprise that needs maximize its profit. The Profit function is: $$\pi=pf(S,I,N,P)-w_1S-w_2I-w_3P-w_4N$$
Profit maximization is achieved when:  \\ \\(1)  \(p\frac{\partial f}{\partial S}=w_1\)  (2)  \(p\frac{\partial f}{\partial I}=w_2\)   (3)  \(p\frac{\partial f}{\partial P}=w_3\)  (4)  \(p\frac{\partial f}{\partial N}=w_4\)\\ \\

Performing calculations the following values of S, I, P and N are obtained:

\begin{eqnarray}\nonumber
S & = &\left(pk\alpha^{1-\left(\beta+\gamma+\delta\right)}\beta^{\beta}\gamma^{\gamma}\delta^{\delta} \right.\\ 
&& \left. w_1^{\beta+\gamma +\delta-1}w_2^{-\beta}w_3^{-\gamma}w_4^{-\delta}\right)^\frac{1}{1-\left(\alpha+\beta+\gamma+\delta\right)}
\end{eqnarray}

\begin{eqnarray}\nonumber
I&=&\left(pk\alpha^{\alpha}\beta^{1-\left(\alpha+\gamma+\delta\right)}\gamma^{\gamma}\delta^{\delta}\right.\\
&& \left.w_1^{-\alpha}w_2^{\alpha+\gamma +\delta-1}w_3^{-\gamma}w_4^{-\delta}\right)^\frac{1}{1-\left(\alpha+\beta+\gamma+\delta\right)}
\end{eqnarray}

\begin{eqnarray}\nonumber
P&=&\left(pk\alpha^{\alpha}\beta^{\beta}\gamma^{1-\left(\alpha+\beta+\delta\right)}\delta^{\delta}\right.\\
&& \left.w_1^{-\alpha}w_2^{-\beta}w_3^{\alpha+\beta +\delta-1}w_4^{-\delta}\right)^\frac{1}{1-\left(\alpha+\beta+\gamma+\delta\right)}
\end{eqnarray}

\begin{eqnarray}\nonumber
N&=&\left(pk\alpha^{\alpha}\beta^{\beta}\gamma^{\gamma}\delta^{1-\left(\alpha+\beta+\gamma\right)}\right.\\
&& \left.w_1^{-\alpha}w_2^{-\beta}w_3^{-\gamma}w_4^{\alpha+\beta +\gamma-1}\right)^\frac{1}{1-\left(\alpha+\beta+\gamma+\delta\right)}
\end{eqnarray}

which is the equation for  data center's profit maximizing quantity of output, as a function of prices of output and inputs.\\ \(y\) increases in price of its output and decreases in price of its inputs iff :
\begin{equation}1-(\alpha+\beta+\gamma+\delta)>0\nonumber\end{equation}
\begin{equation}
 \alpha+\beta+\gamma+\delta<1\nonumber\end{equation}Therefore, the enterprise will have profit maximization at the phase of decreasing returns to scale. It is shown in \cite{[10]}, that profit maximization is scalable i.e. for an arbitrary $n$, number of input variables(constant), the result stands as long as \(\sum_{i=1}^{n}\alpha_i<1\);   where \(\alpha_i\) is the \(i^{th}\) elasticity of the input variable \(x_i\). Constructing a mathematical model using Cobb-Douglas Production Function helps in achieving the following goals:
	
\begin{enumerate}
\item To forecast the revenue with a given amount of investment or input cost.
\item Analysis of maximum production such that total cost does not exceed a particular amount.
\item Analysis of maximum profit that can be achieved.
\item Analysis of minimum cost /input to obtain a certain output.
\end{enumerate}

The model empowers the IaaS entrepreneurs (while establishing an IT data center) estimate the probable output, revenue and profit. It is directly related to a given amount of budget and its optimization. Thus, it deals with minimization of costs and maximization of profits too.\\
The assumption of those 4 factors (S,I,P,N) as the inputs relevant to the output of an IaaS data center is consistent with the work-flow of such data centers. Again, \(\alpha\), \(\beta\), \(\gamma\) and \(\delta\) are assumed to be output elasticity of servers, infrastructure, power drawn and network respectively. A quick run down of the analytical work reveals the application of the method of Least Squares by meticulously following all the necessary mathematical operations such as making the Production Function linear by taking log of both sides and applying Lagrange Multiplier and computing Maxima/Minima by partial differentiation (i.e., computing changes in output corresponding to infinitesimal changes in each input by turn). In a nutshell, the analytical calculation of Marginal Productivity of each of the 4 inputs has been performed. Based on the construction, the mathematical model is capable of forecasting output, revenue and profit for an IaaS data centre, albeit with a given amount of resource or budget. 
\subsection{Observations}
Does this model anyway contradict the established laws of neo-classical economics anyway? \\

Neo-classical economics at abstract level, postulates Average Cost Curve (AC) to be a U-shaped curve whose downward part depicts operation of increasing returns and upward the diminishing returns. Actually, it is the same phenomenon described from two different perspectives; additional applications of one or two inputs while others remaining constant, resulting in increase  in output but at a diminishing rate or increase in marginal cost (MC) and concomitantly  average cost (AC).
\begin{figure}[h]
  \includegraphics[width=80mm]{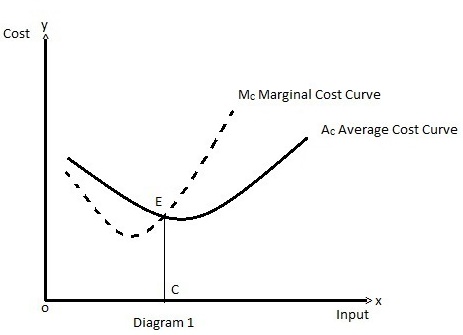}
  \caption{Input vs Cost-1}
  \label{fig:Input vs Cost-1}
\end{figure}

Figure-1 shows that the cost is lowest or optimum where MC intersects AC at its bottom and then goes upward.
 
 \begin{figure}[h]
  \includegraphics[width=80mm]{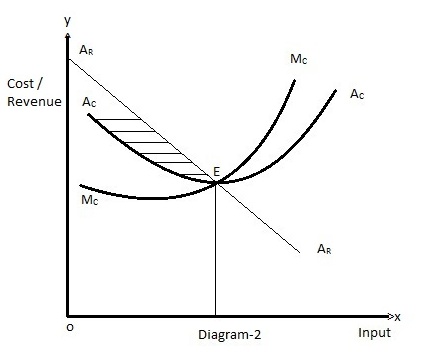}
  \caption{Input vs Cost-2}
  \label{fig:Input vs Cost-2}
\end{figure}

Figure-2 shows that equilibrium (E) or maximum profit is achieved at the point where Average Revenue Curve ( a left to right downward curve, also known as Demand Curve or Price Line as it shows gradual lowering of marginal and average revenue intersects (equals) AC and MC at its bottom, i.e., where AR=AC=MC.
 Here, the region on the left of point E, where AR $>$ AC depicts total profit. Therefore, E is the point of maximization of profit where AR = AC.\\

The Data Center Comparison Cost table data [12] has been accumulated from the figure 3. 

\begin{figure}[h]
  \includegraphics[width=90mm]{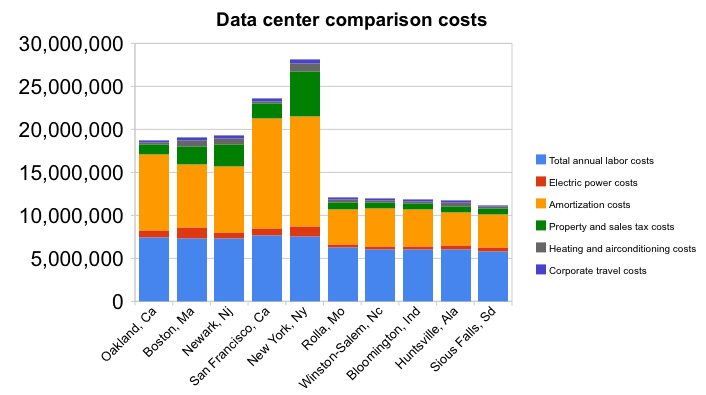}
  \caption{Data Center Comparison Cost}
  \label{fig:Data Center Comparison Cost}
\end{figure}

Additional data files uploaded to gitHub, an open repository, \cite{[12]} documents detailed costs associated with different data centers located in different cities. Along with that, the maximum revenue, which is achievable using the Cobb-Douglas function, is shown. The optimal values of the elasticity constants are also visible in two columns. Additional files contain the proof of scalability of our model. \\
We have partitioned all the segments of Data Center costs into two portions. Considering Labor, property sales tax, Electric power cost as infrastructure and combining Amortization, heating air-conditioning as recurring, we have recalculated the costs of all the data centers. The cost of running data center in New York is highest as its annual labor cost, sales and power costs are higher than any other cities. The operating costs of data center in cities such as Rolla, Winston-Salem, and Bloomington are ranging within \$11,000,000 to 12,500,000, are almost equal.

\begin{figure}[h]
  \includegraphics[width=100mm]{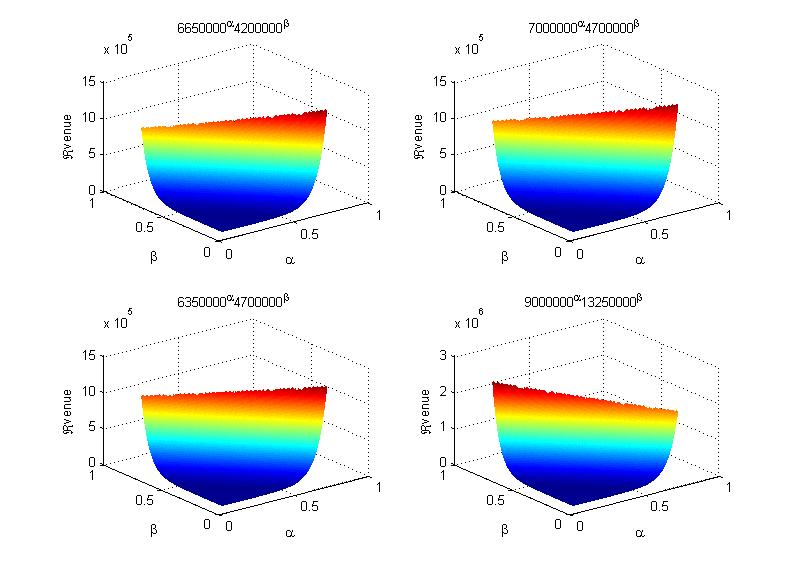}
  \caption{Revenue Function Graph of Data Center Comparison Cost of DRS}
  \label{fig:Revenue Function Graph of Data Center Comparison Cost of DRS}
\end{figure}

In figure 4, X axis represents \(\alpha\); Y axis presents \(\beta\) and Z axis displays the Revenue. The graph demonstrates an example of concave graph. \(\alpha\) and \(\beta\) are the output elasticity of infrastructure and recurring costs. The recurring and infrastructure costs of data centers located in Huntsville-Ala, Rolla-Mo, Bloomington-Ind, and San Francisco-Ca have been used to plot in the above graphs. We can see the revenue is healthier where \(\alpha\), \(\beta\) are both higher than 0.7. The max revenue is lying in the region where \(\alpha\), \(\beta\) are proximal to 0.8 and 0.1 respectively. We choose these values! This selection is verified later by deploying Least Square like fitting algorithms, as discussed in Section \textbf{VI}.

\begin{figure}[h]
  \includegraphics[width=90mm]{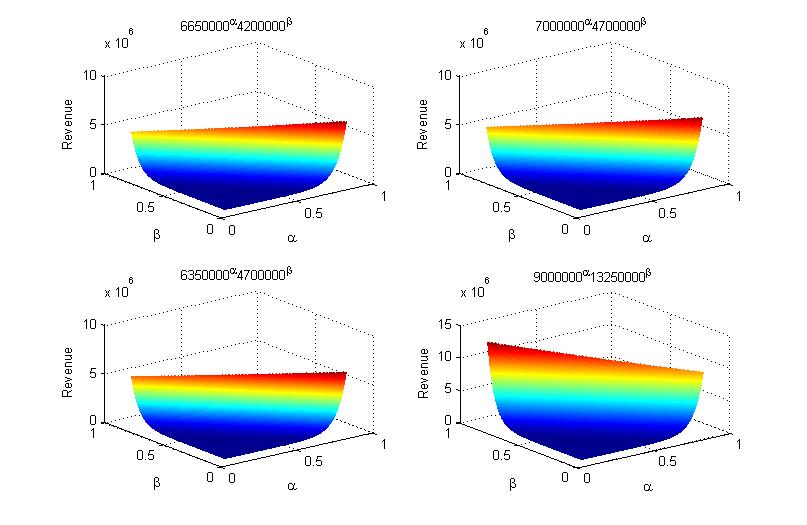}
  \caption{Revenue Function Graph of Data Center Comparison Cost of CRS}
\label{fig:Revenueunction_graph_of_Data_Center_Comparison_Cost}
\end{figure}

The graphs (Figure 5) portray the effects of Cobb-Douglas production function over cost incurred in different data centers located in different cities. As par pictorial representation, there is not much difference with DRS though the revenues obtained in CRS are higher in comparison to DRS. The observation is visible through the data available in table. Similar to the other graphs, the X, Y, Z axes represent \(\alpha\), \(\beta\) and Revenue respectively.\\


\begin{figure}[h]
  \includegraphics[width=90mm]{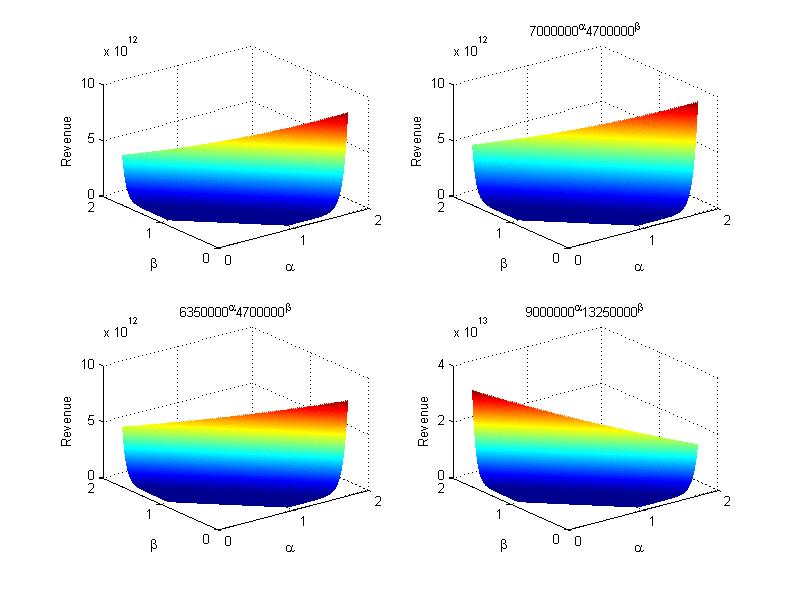}
\caption{Revenue Function Graph of Data Center Comparison Cost of IRS}
\label{fig:5Revenue_function_graph_of_Data_Center_Comparison_Cost}
\end{figure}

Figure 6 depicts the revenue under the constraint, Increasing return to scale (IRS)  where the sum of the elasticities is more than 1. Like the previous figures, the elasticities are represented by the X, Y axes and Z represents the revenue, which has been calculated using Cobb-Douglas function.\\
Additional file \cite{[12]} contains detailed information about data center comparison costs for IRS,DRS and CRS, including revenue data, cost and optimal constraints. Please refer \cite{[10]} for a quick tutorial on IRS, DRS and CRS.


\begin{figure}[h]
  \includegraphics[width=90mm]{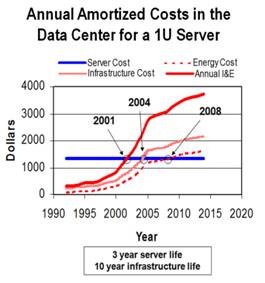}
\caption{Annual Amortization Costs in data center for 1U server}
\label{fig:6 Annual Amortization Costs in data center for 1U server}
\end{figure}

Figure 7 is the graphical representation of Annual Amortization Costs in data center for 1U server. All units are in \$. We have extracted fairly accurate data from the graph and represented in tabular format (Table IX). Maximum revenue and optimal elasticity constants are displayed in the same table.

[12] shows the  Optimal constants for DRS.
\begin{figure}[h]
  \includegraphics[width=90mm]{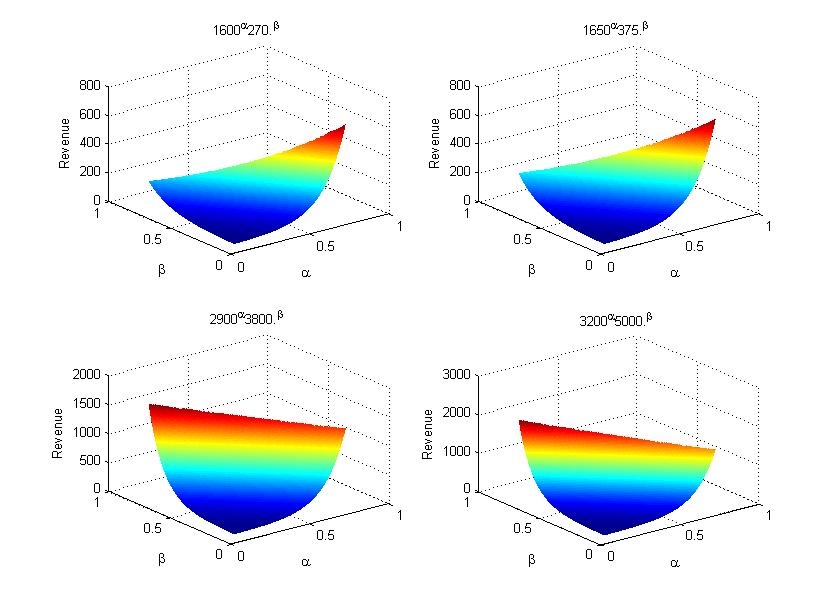}
\caption{Revenue Function Graph of Annual Amortized Cost}
\label{fig:7Revenue_function_Graph_of_Annual_Amortized_Cost}
\end{figure}

The Revenue graph (Figure 8) displays the range of revenue in accordance to the data of annual amortized cost of different years.The co-ordinate axes represent \(\alpha\), \(\beta\) and Revenue respectively. Server cost and Infrastructure cost are combined together as infrastructure cost, whereas Energy and Annual I \& E are clubbed as recurring cost.  \(\alpha\) represents elasticity constant of infrastructure and \(\beta\) denotes elasticity constant of recurring cost. The recurring cost and infrastructure cost of the years 1992, 1995, 2005, and 2010 have been used to calculate revenue. The revenue rises drastically in region of \(\alpha\), \(\beta\)  being greater than 0.5 in comparison to any other region. The peak of the graphs indicate the maximum revenue located in the region, where \(\alpha\), \(\beta\) are approximating 0.8 and 0.1 or vice-versa.

\begin{figure}[h]
  \includegraphics[width=90mm]{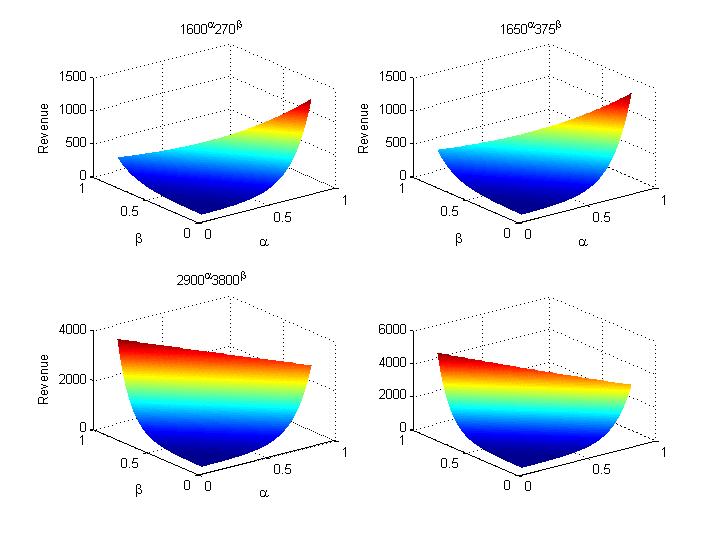}
\caption{Revenue Function Graph of Annual Amortized Cost}
\label{fig:8Revenue_function_Graph_of_Annual_Amortized_Cost}
\end{figure}

In the Figure 9, the co-ordinate axes represent \(\alpha\), \(\beta\), and revenue respectively. Slight difference is observed in the range of elasticities. Maximum revenue lies in the area, where (\(\alpha\) is approximately 0.9 and \(\beta\) is close to 0.1 or vice versa. The revenue data, elasticities and different cost segments are displayed in tabular format[11].

We observe that there is no major difference between revenues during the years 1992 to 1995. But the revenue becomes almost 3 fold between the years 2000 and 2010. Server cost remains constant throughout the years but significant changes are noticed in other cost segments namely Energy cost, Infrastructure and Annual I \& E.

\begin{figure}[h]
  \includegraphics[width=90mm]{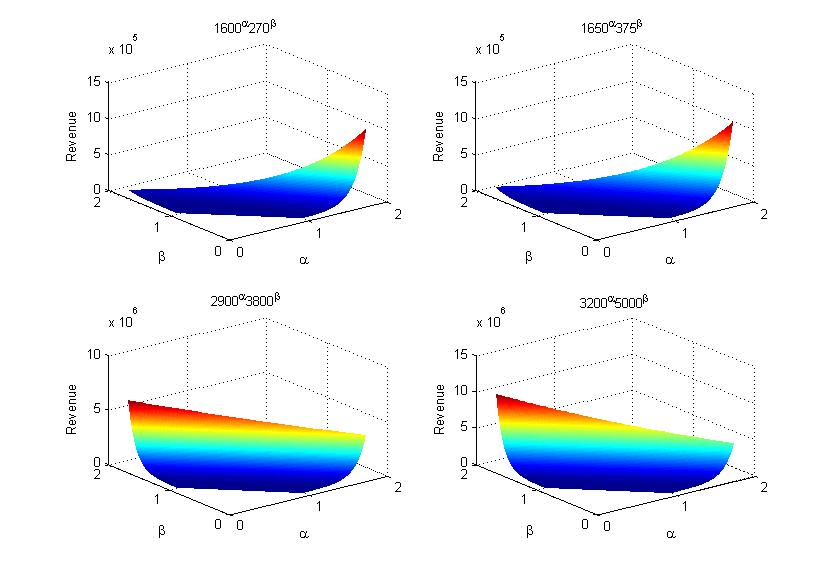}
  
\caption{Revenue Function Graph of Annual Amortized Cost}
\label{fig:pRevenue_function_Graph_of_Annual_Amortized_Cost}
\end{figure}

In Figure 10, the maximum revenue is reflected in the region where (\(\alpha\) and \(\beta\) nearby 1.8 and 0.1 respectively. In case of IRS, the optimal revenue surges ahead of CRS and DRS. Revenue becomes almost five times from the year 2000 to 2010. It displays two-fold jump from 2005 to 2010.

\section{Design of Experiments (DoE) and impact on the proposed model} \label{sec:factor}
Factor analysis is an efficient way to understand and analyse data. Factor analysis contains two types of variables, latent and manifest. A DoE paradigm identifies  the latent(unobserved) variable  as a function of manifest(observed) variable. Some well known methods are  principal axis factor, maximum likelihood, generalized least squares, unweighted least square etc. The advantage of using  factor analysis is to identify the similarity between manifest variables and latent variable. The number of factors correspond to the variables. Every factor identifies the impact of overall variation in the observed variables. These factors are sorted in the order of variation they contributed to overall result. The factors which have significantly lesser contribution comapared to the dominat ones may be discarded without causing significant change in output. The filteration process is rigorous but the outcome is insightful.

\subsection{Factor Analysis for Proposed model}
The proposed revenue model exploits factorial analysis to identify the major factors among the inputs(Cost of Servers and Power). Table I, Table II and Table III have been created from the data available. Equation (13) describes our basic model with two factors each with two levels, defining all combinations for the output variable. Factorial design identifies the percentage of contribution of each factor. These details can be used to understand and decide how the factors can be controlled to generate better revenue. 
The Cobb-Douglas production function provides insight for maximizing the revenue. The paper[10] explains the coefficient of the latent variables as a contributor of output function as evident from equation (2). In the given equation, (\(\alpha\) , \(\beta\) , \(\gamma\) and \(\delta\) are the parameters which are responsible for controlling the output function Y. However to generate the output function y=f(S,I,N,P), the threshold level of minimum and maximum value needs to be bounded. The contribution of (\(\alpha\) , \(\beta\) , \(\gamma\) and \(\delta\) towards output function Y is not abundantly clear.  \\
Therefore, it is relevant to study the effects of such input variables on the revenue in terms of percentage contribution of each variable. An efficient, discrete factorial design is implemented to 
study the effects and changes in all relevant parameters regarding revenue. Revenue is modeled depending on a constant(market force), a bunch of input variables  which are quantitative or  categorical in nature. The road-map to design a proper set of experiments for simulation involves the following:

\begin{itemize}
\item Develop a model best suited for the data obtained.
\item Isolate measurement errors and gauge confidence intervals for model parameters.
\item Ascertain the adequacy of the model.
\end{itemize}  
 
For the sake of simplicity and convenience the factors S-I and P-N were grouped together as two factors. The question of scaling down impacting the model performance would be asked is not the limitation of the model. Additional files, \cite{[12]} reveal a proof which considers $n$ number of factors for the same model and the conditions for optimality hold,$n$ being arbitrary . Additionally, the conditions observed for two factors can be simply scaled to condition for n factors. Since we consider the equation with two factors only,the equation can be rewritten as

\begin{equation} f(S,P)=AS^\alpha P^\beta
\end{equation}
\begin{equation}     =AS^\alpha P^(1-\alpha)
\end{equation}
 For A=1
 Profit maximization is achieved when:
 \\ \\(1)  \(\frac{\partial y}{\partial S}\)=\(\frac{\alpha S^(\alpha -1) K^(1 -\alpha)}{k}\)

  \hspace{.6cm}= \(\frac{\alpha Y}{k}\)
  
  Profit maximization is achieved when:
\\ \\(2)  \(\frac{\partial y}{\partial K}\)=\(\frac{(1-\alpha) S^(\alpha) K^(1 -\alpha)}{k}\) 

  \hspace{.7cm}= \(\frac{(1-\alpha) Y}{k}\)
  
  At this point, we note that both the factors,servers and power can be controlled by alpha. The rate of change in both the parameters in the Cobb-Douglas equation can be determined. We have to choose the alpha value in such a way that the profit maximization would not drop below the threshold value.

\subsection *{\textbf{The \(2^2\) factorial design}}

The following are the factors, with 2 levels each:

\begin{enumerate}
\item Cost of Power and Cooling, Factor $1$ .
\item Type of Server categorized based on cost of deployment, Factor $2$. 
\end{enumerate}
\begin{table}[h]
\centering
\begin{tabular}{|c|c|}
\hline
Level & Range (in Million Dollars) \\
\hline
Low & 5-15\\
\hline
High & 16-40\\
\hline
\end{tabular}
\caption{Factor 1 Levels}
\end{table}

\begin{table}[h]
\centering
\begin{tabular}{|c|c|}
\hline
Level & Range (in Million Dollars) \\
\hline
Type 1 & 45-55\\
\hline
Type 2 & 56-65\\
\hline
\end{tabular}
\caption{Factor 2 Levels}
\end{table}

Let us define two variables \(x_A\) and \(x_B\) as: 
\begin{equation}
    x_A=
    \begin{cases}
      -1, & \text{if Factor 1 is low}  \\
       1, & \text{if Factor 1 is high}\\
    \end{cases}
        \nonumber
  \end{equation}
  \begin{equation}
    x_B=
    \begin{cases}
      -1, & \text{if Factor 2 is of Type-1}  \\
       1, & \text{if Factor 2 is of Type-2}\\ 
    \end{cases}
        \nonumber
  \end{equation}
  The Revenue y (in Million Dollars) can now be regressed on \(x_A\) and \(x_B\) using a nonlinear regression model of the form:
  \begin{equation} y = q_0+q_Ax_A+q_Bx_B+q_{AB}x_Ax_B\end{equation}
  The effect of the factors is measured by the proportion of total variation explained in the response.\\

Sum of squares total (SST):
\begin{equation} 
SST=2^2{q_A}^2+2^2{q_B}^2+2^2{q_{AB}}^2
\end{equation}
Where:\\\(2^2\)\(q^2
_A\) is the portion of SST that is explained by Factor 1.\\
\(2^2\)\(q^2_B\) is the portion of SST that is explained by Factor 2.\\
\(2^2\)\(q^2_{AB}\) is the portion of SST that is explained by the interactions of Factor 1 and Factor 2.\\
Thus, \begin{equation}\text{SST = SSA + SSB + SSAB}\end{equation}
\begin{equation}\text{Fraction of variation explained by A}= \frac{SSA}{SST}
\end{equation}

\begin{equation}\text{Fraction of variation explained by B}= \frac{SSB}{SST}
\end{equation}

\begin{equation}\text{Fraction of variation explained by AB}= \frac{SSAB}{SST}
\end{equation}

Our choice of elasiticity depends on the dynamics between the factors and the benchmark constraints of optimization.				
CRS, for example requires the sum of the elasticities to equal 1 and DoE reveals that factor1 contributes to the response variable to a lesser extent compared to factor 2.				
Therefore, in order that revenue growth may be modeled in a balanced fashion, elasticity value for factor 1 has been set to  much higher compared to factor 2.	
The same phenomena is observed in the cases of IRS and DRS and identical heuristic has been applied to pre-determine the choice of elasticities.				
The authors intend to verify the heuristic choices through fitting and regression in the latter part of the manuscript.		

\begin{table}[!htpb]
\centering
 \begin{tabular}{|c|c|c|c|c|}
\hline 
 Read Data	& Factor 1 &	Factor 2 &	Alpha & 	Beta\\
 \hline				
CRS	& 24.5 &	75.8	& 0.9 &	0.1\\
\hline	
DRS	& 2.44	& 97.36 &	0.8 &	0.1\\
\hline	
IRS	& 5.86	& 93.62 &	1.8 &	0.1\\
\hline	

\end{tabular}
\caption{Elasticity and percentage contribution of cost factors}
\end{table}


\section{Experimental Observations}
\label{S:2}
\subsection{Experiment 1 : IRS }
\begin{table}[h]
\centering
 \begin{tabular}{|c|c|c|}
\hline 
  & \multicolumn{2}{c|}{Power and Cooling} \\
\hline 
  Server & Low & High\\
  \hline
  Type-1 & 1509.63 & 1676.48\\
  \hline
  Type-2 & 2062.39 & 2153.34\\
  \hline
\end{tabular}
\caption{Experiment 1: IRS}
\end{table}

\underline{Computation of Effects}
\\ \\ \\Substituting the four observations in the model, we obtain the following equations:\\

\begin{equation}1509.63 = q_0-q_A-q_B+q_{AB}\nonumber\end{equation}
\begin{equation}1676.48 = q_0+q_A-q_B-q_{AB}\nonumber\end{equation}
\begin{equation}2062.39 = q_0-q_A+q_B-q_{AB}\nonumber\end{equation}
\begin{equation}2153.34 = q_0+q_A+q_B+q_{AB}\nonumber\end{equation}
\\

Solving the above equations for the four unknowns, the Regression equation obtained is:
\begin{equation}y = 1850.46 + 64.45 x_A + 257.4 x_B - 18.9 x_Ax_B\end{equation}
If we spend on a server for deployment and capacity building ,revenue is positively affected. Cloud business elasticity depends on resources and capacity and promise of elastic service provisioning is a function of hardware and software capabilty . 

\underline{Allocation of Variation}:
\begin{equation} 
\begin{split}
\text{SST} & = 2^2{q_A}^2+2^2{q_B}^2+2^2{q_{AB}}^2\\
 & = 2^2{64.45}^2+2^2{257.4}^2+2^2{-18.9}^2\\
 & = 283063\nonumber
\end{split}
\end{equation}
The result is interpreted as follows:\\
The effect of Factor 1 on Revenue is 5.86\%\\
The effect of Factor 2 on Revenue is 93.62\%\\
The effect of interactions of Factors 1 and 2 on Revenue is 0.5\%\\

\subsection{Experiment 2: CRS }

\begin{table}[h]
\centering
 \begin{tabular}{|c|c|c|}
\hline 
  & \multicolumn{2}{c|}{Power and Cooling} \\
\hline 
  Server & Low & High\\
  \hline
  Type-1 & 43.5 & 47.5\\
  \hline
  Type-2 & 50.82 & 55.38\\
  \hline
\end{tabular}
\caption{Experiment 2: CRS}
\end{table}
The regression equation obtained is:
\begin{equation}y = 49.3 + 2.14 x_A + 3.8 x_B + 0.14 x_Ax_B\end{equation}
The result obtained after factor analysis is interpreted  as follows:\\
The effect of Factor 1 on Revenue is 24.5\%\\
The effect of Factor 2 on Revenue is 75.8\%\\
The effect of interactions of Factors 1 and 2 on Revenue is 0.1\%\\

\subsection{Experiment 3: DRS }

\begin{table}[h]
\centering
 \begin{tabular}{|c|c|c|}
\hline 
  & \multicolumn{2}{c|}{Power and Cooling} \\
\hline 
  Server & Low & High\\
  \hline
  Type-1 & 29.47 & 31.99\\
  \hline
  Type-2 & 33.6 & 36.87\\
  \hline
\end{tabular}
\caption{Experiment 3: DRS}
\end{table}

The regression equation is:
\begin{equation}y = 41.9 + 1.77 x_A + 11.18 x_B + 0.5 x_Ax_B\end{equation}
The result obtained after factor analysis is interpreted as follows:\\
The effect of Factor 1 on Revenue is 2.44\%\\
The effect of Factor 2 on Revenue is 97.36\%\\
The effect of interactions of Factors 1 and 2 on Revenue is 0.19\%\\

The results suggest that there is no significant interaction between the factors. Thus, all further analysis henceforth will be done ignoring the interaction factor.

\subsection{Randomizing the data}
Since there was insufficient data to conclude the effects of the factors on revenue, we had to generate more data by discovering the distribution of the real data set and generating random data following the same distribution. Our experiment has found that the original data follows the Normal distribution (Figure 11 and 12).

\begin{figure}[!htb]
  \includegraphics[width=90mm]{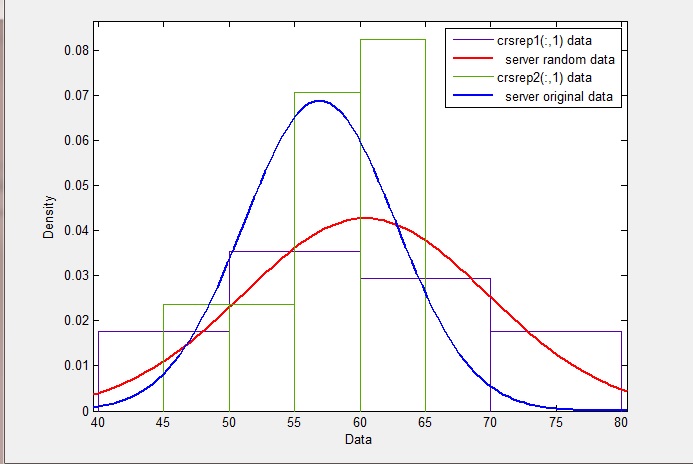}
  \caption{The Original and Generated Server Data that follows Normal Distribution}
  \label{The Original and Generated Server Data that follows Normal Distribution}
\end{figure}

\begin{figure}[!htpb]
  \includegraphics[width=90mm]{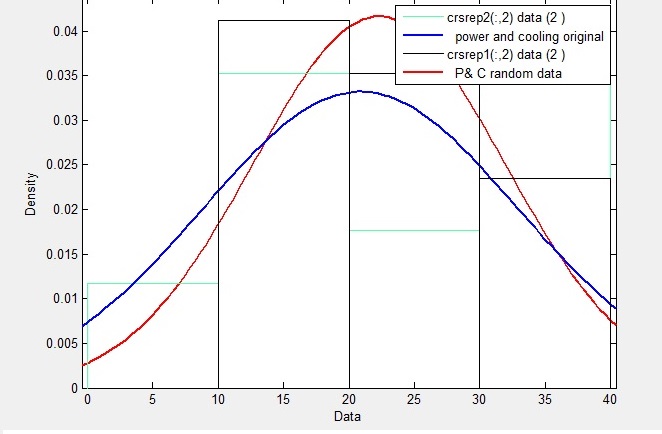}
  \caption{The Original and Generated Power and Cooling Data that follows Normal Distribution}
  \label{The Original and Generated Power and Cooling Data that follows Normal Distribution}
\end{figure}


The tables VI,VII represent the random data that was generated and corresponding revenue values calculated using the Cobb-Douglas model for IRS, CRS and DRS respectively.

The Chi Square- Goodness of fit test was performed on the actual and generated data to confirma the data trend.

The Null Hypotheses \(H_0\): After adding noise to the original data set, the data follows Gaussian Distribution.
If \(H_0\)=1, the null hypothesis is rejected at 5\% significance level.

if \(H_0\)=0, the null hypothesis is accepted at 5\% significance level.

The result obtained, \(H_0\)=0 assures us that the data indeed follows Gaussian Distribution with 95\% confidence level.

\begin{table}[!htb]
\centering
 \begin{tabular}{|c|c|c|}
\hline 
 New Server & Power and Cooling & Revenue\\
\hline 
68 & 38 & 2860.85	\\
\hline
44 & 25 & 1253.17\\
\hline
62 & 12 & 2158.85\\
\hline
59 & 37 & 2209.80\\
\hline
49 & 10 & 1387.88	\\
\hline
54 & 20 & 1771.79	\\
\hline
59 & 18 & 2056.18	\\
\hline
78 & 25 & 3512.08	\\
\hline
73 & 25 & 3117.28	\\
\hline
49 & 10 & 1387.88	\\
\hline
75 & 21 & 3216.12	\\
\hline
61 & 19 & 2195.17\\
\hline
57 & 28 & 2019.72	\\
\hline
61 &34 & 2326.71\\
\hline
56 &34 & 1994.74	\\
\hline
56 & 11 & 1781.88	\\
\hline
66 & 12 & 2566.97	\\
\hline
\end{tabular}
\caption{Revenue for IRS}
\end{table}

\begin{table}[!htb]
\centering
 \begin{tabular}{|c|c|c|}
\hline 
 New Server & Power and Cooling & Revenue\\
\hline 
68	 & 38	& 64.16\\
\hline 
44	&  25	& 41.58\\
\hline 
62	 & 12	& 52.61\\
\hline 
59	& 37	&  56.31\\
\hline 
49	 & 10	&  41.80\\
\hline 
54	 & 20	&  48.89\\
\hline 
59	&  18	& 52.40\\
\hline 
78	& 25	& 69.61\\
\hline 
73	& 25	 & 65.58\\
\hline 
49	& 10	& 41.80\\
\hline 
75	 & 21	& 66.04\\
\hline 
61	& 19	& 54.28\\
\hline 
57	& 28	& 53.09\\
\hline 
61	& 34	& 57.54\\
\hline 
56	& 34	& 53.27\\
\hline 
56 & 11	& 47.59\\
\hline 
66	& 12	& 55.66\\
\hline
\end{tabular}
\caption{Revenue for CRS}
\end{table}

\begin{table}[!htb]
\centering
 \begin{tabular}{|c|c|c|}
\hline 
 New Server & Power and Cooling & Revenue\\
\hline 
68	& 38	& 42.0713\\
\hline
44	& 25	& 28.4811\\
\hline
62	& 12	& 34.8202\\
\hline
59	& 37	& 37.4543\\
\hline
49	& 10	& 28.3241\\
\hline
54	& 20	& 32.8109\\
\hline
59	& 18	& 34.8505\\
\hline
78	& 25	& 45.0267\\
\hline
73	& 25	& 42.7024\\
\hline
49	& 10	& 28.3241\\
\hline
75	& 21	& 42.8816\\
\hline
61	& 19	& 35.9865\\
\hline
57	& 28	& 35.4336\\
\hline
61	& 34	& 38.1427\\
\hline
56	& 34	& 35.6204\\
\hline
56	& 11	& 31.8192\\
\hline
66 	& 12	& 38.8935\\
\hline
\end{tabular}
\caption{Revenue for DRS}
\end{table}

\subsection{Replications}

Replicates are multiple experimental runs with identical factor settings (levels). Replicates are subject to the same sources of variability, independent of each other. 
 
In the experiment, two replications were conducted on the real data and generated data(r=2), taking into consideration that it is a \(2^2\) factorial design problem. 
The results obtained are at par with the results obtained from factorial analysis conducted for the original data.
Replication, the repetition of an experiment on a large group of subjects, is required to improve the significance of an experimental result. If a treatment is truly effective, the long-term averaging effect of replication will reflect its experimental worth. If it is not, then the few members of the experimental population who may have reacted to the treatment will be negated by the large numbers of subjects who were unaffected by it. Replication reduces variability in experimental results, increasing their significance and the confidence level with which a researcher can draw conclusions about an experimental factor \cite{[13]} . Since this was a \(2^2\) Factorial problem, 2 replications had to be performed. Table IX documents the results obtained after replications were performed for IRS, CRS and DRS respectively.

\begin{table}[!htpb]
\centering
 \begin{tabular}{|c|c|c|c|}
\hline 
& IRS & CRS & DRS\\
\hline
New Server &81.9 & 66.21 &62.19 \\
\hline
Power and cooling &12.48 & 31.43 &35.72\\
\hline
Interaction &3.05 & .35 & 0.000013\\
\hline
Error & 2.55 & 1.99& 2.02\\
\hline
\end{tabular}
\caption{Percentage variation for IRS, CRS and DRS}
\end{table}

It is observed that the contribution of two factors towards the total variation of the response variable is consistent between the real data and the simulated random data.

\subsection{Confidence intervals for effects}
The effects computed from a sample are random variables and would be different if another set of experiments is conducted. The confidence intervals for the effects can be computed if the variance of the sample estimates are known.

If we assume that errors are normally distributed with zero mean and variance \(\sigma_e^2\), then it follows from the
model that the \(y_i\) 's are also normally distributed with the same variance \(\sigma_e\). 

The variance of errors can be estimated from the SSE as follows:

\begin{equation}
s_e^2=\frac{SSE}{2^2(r-1)} \nonumber
\end{equation}

The quantity on the right side of this equation is called the Mean Square of Errors (MSE). The denominator
is $ 2^{2}\left (r - 1\right)$, 
which is the number of independent terms in the SSE. 
\\This is because the r error terms
corresponding to the r replications of an experiment should add up to zero. Thus, only $r-1$ of these terms can
be independently chosen. 

Thus, the SSE has $ 2^{2}\left(r-1\right) $ degrees of freedom.
The estimated variance is $ S_{q0}=S_{qA}=S_{qAB}=\frac{S_{e}}{\sqrt{2^{2}r}}$

The confidence interval for the effects are :
\begin{center}$q_{i}\mp t_{\left[1-\alpha /2;2^{2} \left(r-1\right)\right]^{s_{qi}}}$ \end{center}

The result obtained for the range is as follows:\\
(61.21, 62.15)\\
(-10.58, -9.64)\\
(15.78, 16.72)\\
(-13.36, -12.42)\\

against the actual values 61.68, -10.11, 16.25, -12.89. None of the confidence intervals included  $0$ fortifying the goodness of the experiment.

\subsection{Principal Representative Feature(PRF)} 
The PRF primarily identifies the contributors in the system which has maximum variance and tries to identify a pattern in a given data set which is unique. 

The first principal component accounts for as much of the variability in the data as possible, and each succeeding component accounts for as much of the remaining variability as possible.

Though primarily used for dimensionality reduction, the PRF has been exploited here to figure out the contributions of each factor towards the variation of the response variable. The authors don't intend to ignore one of the two input parameters and that's not how the method should be construed. Since data trends have evidence of normal behavior, PRF was used as an alternative to factor analysis. If Shapiro Wills test for normalcy revealed non-normal behavior, ICA could have been used to understand how each factor contributes to the response variable, "y".

The PRF conducted on the generated data gave the following results:
variation explained by first factor (New server) is 66\%
2nd factor (P\&C) explains 34.08\% of the variation. 

\subsection{Non-parametric Estimation}
   A parametric statistical test is one that makes assumptions about the parameters (defining properties) of the population distribution(s) from which one's data are drawn, while a non-parametric test is one that makes no such assumptions.
   
    The tests involve estimation of the key parameters of that distribution ( the mean or difference in means) from the sample data. The cost of fewer assumptions is that non-parametric tests are generally less powerful than their parametric counterparts.
    
  Apart from the conclusion obtained above, we perform the non-parametric estimation which does not rely on assumptions that the data are drawn from a given probability distribution.

  \begin{figure}[!htpb]
  \includegraphics[width=90mm]{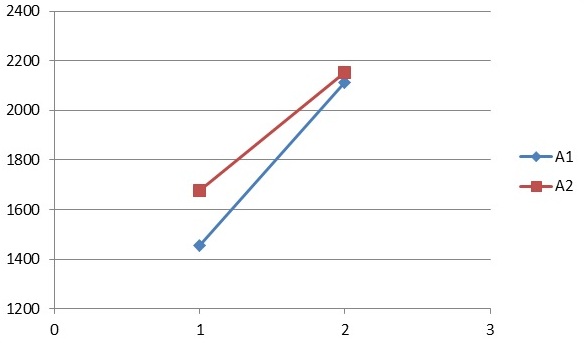}
  \caption{Non-parametric Estimation for IRS-Original Data}
  \label{Non-parametric Estimation for IRS-Original Data}
\end{figure}

\begin{figure}[!htpb]
  \includegraphics[width=90mm]{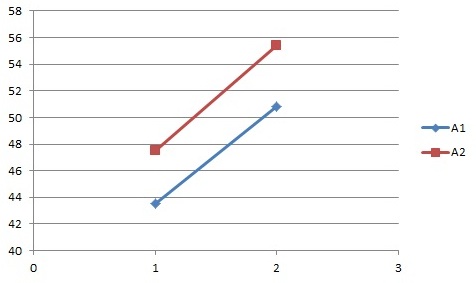}
  \caption{Non-parametric Estimation for CRS-Original Data}
  \label{Non-parametric Estimation for CRS- Original Data}
\end{figure}

\begin{figure}[!htpb]
  \includegraphics[width=90mm]{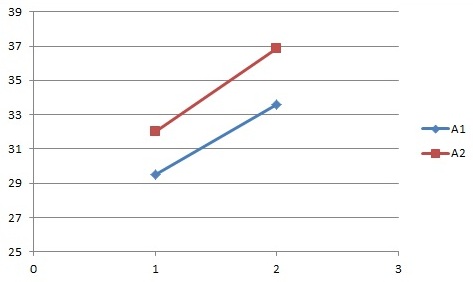}
  \caption{Non-parametric Estimation for DRS-Original Data}
  \label{Non-parametric Estimation for DRS-Original Data}
\end{figure}
\begin{figure}[!htpb]
  \includegraphics[width=90mm]{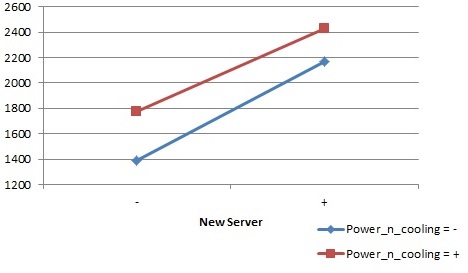}
  \caption{Non-parametric Estimation for IRS-Generated Data}
  \label{Non-parametric Estimation for IRS-Generated Data}
\end{figure}

\begin{figure}[!htpb]
  \includegraphics[width=90mm]{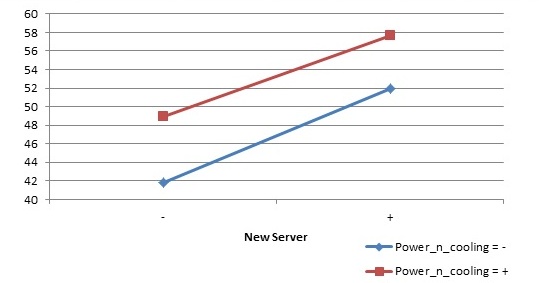}
  \caption{Non-parametric Estimation for CRS-Generated  Data}
  \label{Non-parametric Estimation for CRS-Generated Data}
\end{figure}

\begin{figure}[!htpb]
  \includegraphics[width=90mm]{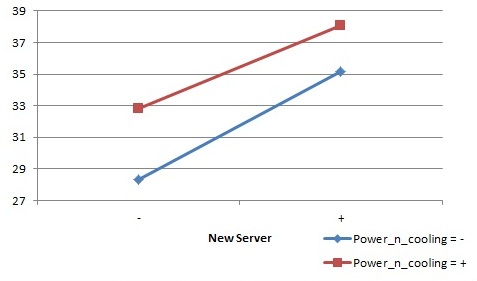}
  \caption{Non-parametric Estimation for DRS-Generated Data}
  \label{Non-parametric Estimation for DRS-Generated Data}
\end{figure}

The figures 13, 14 and 15 represent the results obtained from the non- parametric estimation for the data A1: New Server and A2: Power \& Cooling, corresponding to IRS, CRS and DRS respectively and also visualizes the interaction between the factors.

The figures 16, 17 and 18 represent the results obtained from the non- parametric estimation for the data New Server and Power \& Cooling, corresponding to IRS, CRS and DRS respectively on the generated data set and also visualizes the interaction between the factors.

The above figures suggest no interaction between the factors, which is in agreement with the results obtained in the previous sections.

\section{Experiments}

Let the assumed paramteric form be y = K′ + \(\alpha\) log(S) + \(\beta\) log(P). Consider a set of data points.

\[\begin{array}{ccccccc}
lny_1 & = & K' &{}+{} & \alpha S'_1  & {}+{} & \beta P'_1  \\
 \vdots & & \vdots & & \vdots & & \vdots \\
lny_N & = & K' &{} +{} & \alpha S'_N & {}+{} & \beta P'_N  \\
\end{array}\]

where 
\begin{equation}
S'_i = log(S'_i) \nonumber
\end{equation}
\begin{equation}
P'_i = log(P'_i) \nonumber
\end{equation}

If N \textgreater 3, (23) is an over-determined system. One possibility is a least squares solution. Additionally if there are constraints on the variables (the parameters to be solved for), this can be posed as a constrained optimization problem. These two cases are discussed below.
\begin{enumerate}
\item No constraints : An ordinary least squares solution. (24) is in the form y = Ax where,

$$ x =
 {\begin{bmatrix}
    K' & \alpha & \beta
  \end{bmatrix}}^T
 $$

\begin{equation}
y = \begin{bmatrix}
    y_1\\ . \\ . \\ y_N
  \end{bmatrix}
\end{equation}

and

\begin{equation}
A = \begin{bmatrix}
    1 & S'_1 & P'_1\\ 
      & ...  &     \\ 
    1 & S'_N & P'_N
  \end{bmatrix}
\end{equation}

The least squares solution for x is the solution that minimizes 
\begin{equation}
(y - Ax)^T (y - Ax) \nonumber 
\end{equation} 
It is well known that the least squares solution to (24) is the solution to the system 
\begin{equation}
A^T y= A^TAx \nonumber 
\end{equation}
i.e.
\begin{equation}
x = (A^TA)^{-1} A^T y \nonumber 
\end{equation}

In Matlab the least squares solution to the overdetermined system y = Ax can be obtained by
x = A $\backslash$ y.\\
The following is the result obtained for the elasticity values after performing the least square fitting: 
 
 \begin{table}[!htpb]
\centering
 \begin{tabular}{|c|c|c|c|}
\hline 
& IRS & CRS & DRS\\
\hline
\(\alpha\) & 1.799998 & 0.900000 & 0.799998\\
\hline
\(\beta\) & 0.100001 &  0.100000 & .099999\\
\hline
\end{tabular}
\caption{Least square test results}
\end{table}

\item
Constraints on parameters : This results in a constrained optimization problem. The objective function to be minimized (maximized) is still the same namely
\begin{equation}
(y - Ax)^T (y - Ax) \nonumber 
\end{equation} 

This is a quadratic form in x. If the constraints are linear in x, then the resulting constrained optimization
problem is a Quadratic Program (QP). A standard form of a QP is :
\begin{equation}
\text{min  } x^THx +f^Tx
\end{equation}

s.t.

\begin{equation}
Cx \le b \text{   Inequality Constraint}\nonumber
\end{equation}
\begin{equation}
C_{eq}x = b_{eq} \text{   Equality Constraint}\nonumber
\end{equation}

Suppose the constraints are that \(\alpha\)  and \(\beta\) are \textgreater  0 and \(\alpha\)  + \(\beta\)  $\ge$  1. The quadratic program can be written as (neglecting the constant term \(y^T\)y ).

\begin{equation}
\text{min   } x^T(A^TA)x - 2y^TAx
\end{equation}
s.t.
\begin{equation}
 \alpha > 0 \nonumber
\end{equation} 
\begin{equation}
 \beta > 0 \nonumber
\end{equation} 
\begin{equation}
 \alpha + \beta \le 0 \nonumber
\end{equation}

In standard form as given in (29), the objective function can be written as :

\begin{equation}
x^THx + f^Tx
\end{equation}

where

\begin{equation}
H = A^TA \text{  and  } f=-2A^Ty \nonumber
\end{equation}

The inequality constraints can be specified as : 

$$ C =
 \begin{bmatrix}
    0 & -1 & 0\\
    0 &  0 & -1\\
    0 &  1 &  1 
  \end{bmatrix}
 $$

and

$$ b =
 \begin{bmatrix}
    0 \\
    0 \\
    1 
  \end{bmatrix}
 $$
 
 In Matlab, quadratic program can be solved using the function quadprog.
 
The below results were obtained on conducting Quadratic Programming.

\begin{table}[!htpb]
\centering
 \begin{tabular}{|c|c|c|c|}
\hline
 & IRS & CRS & DRS\\
 \hline
 K & 3.1106 & 0 & 0\\
 \hline
 \(\alpha\) & 1.0050 & 0.9000 & 0.8000 \\
 \hline
 \(\beta\) & 0.1424 & 0.1000 & 0.1000\\
\hline
\end{tabular}
\caption{Quadratic Programming results}
\end{table}

\item Using active set:\\

The experiment conducted used active learning technique to learn about active set. This framework is best suited in our case as it can be applied to different performance targets and all types of classifications. The traditional active-set method is divided into two steps- the first focuses on feasibility, while the second focuses on optimality. An advantage of active-set methods is that the methods are well-suited for initial start, where a good estimate of the optimal active set is used to start the algorithm. This is particularly useful in applications where a sequence of quadratic programs is solved. Interior-point methods compute iterates that lie in the interior of the feasible region, rather than on the boundary of the feasible region. The method computes and follows a continuous path to the optimal solution. Hence it is expensive method. Active-set methods differ from interior point methods in that no barrier term is used to ensure that the algorithm remains interior with respect to the inequality constraints. Instead, attempts are made to learn the true active set which we have used .Active set gave best results out of all the three algorithms which suffices our argument. \\
 
 \end{enumerate}
 
\section{Prediction and Forecasting}

Linear regression is an approach for modeling the relationship between a dependent variable y and one or more explanatory variables denoted by x. When one explanatory variable is used, the model is called simple linear regression. When more than one explanatory variable are used to evaluate the dependent variable, the model is called multiple linear regression model.
Applying multiple linear equation model to predict a response variable y as a function of 2 predictor variables  x1,x2 takes the following form:
\begin{equation}
y=b_0+b_1x_1+b_2x_2+e	
\end{equation}
Here, $ {b_0, b_1, b_2} $ are 3 fixed parameters and e is the error term.  \\
 Given a sample,\\ 
 $ {(x_{11},x_{21},y_1),…(x_{1n},x_{2n},y_n)} $ of n observations the model consist of following n equations\\
 
 \begin{equation}
y_1=b_0+b_1x_{11}+b_2x_{21}+e	
\end{equation}
\begin{equation}
y_2=b_0+b_1x_{12}+b_2x_{22}+e	
\end{equation}
\begin{equation}
y_3=b_0+b_1x_{13}+b_2x_{23}+e	
\end{equation}
\begin{equation}
y_n=b_0+b_1x_{1n}+b_2x_{2n}+e	
\end{equation}
So, we have\\

$$
\left(
\begin{array}{c}
y_{1}\\
\vdots\\
y_{n}
\end{array}
\right)
= \left(
\begin{array}{cccc}
1&x_{11}&\cdots&x_{k1}\\
\vdots&\vdots&\ddots&\vdots\\
1&x_{1n}&\cdots&x_{kn}
\end{array}
\right)
\left(
\begin{array}{c}
b_{1}\\
\vdots\\
b_{k}
\end{array}
\right)
+ \left(
\begin{array}{c}
e_{1}\\
\vdots\\
e_{n}
\end{array}
\right)
$$

\begin{equation}
where\hspace{1cm} k={1...17}
\end{equation}

Or in matrix notation:  $ y = Xb + e$ 	\\

Where:
\begin{itemize}
\item $b$ : A column vector with  17 elements are$ {b_0,b_1,...,b_16}$\\
\item $y$: A column vector of n  observed  values  of $ y = {y_1,...,y_n}$\\
\item $X$: An n  row  by  17  column  matrix  whose  $(i,j+1)^{th}$ element $X_{i,j+1}$ is 1  if  j is 0 else $x_{ij}$
\end{itemize}

Parameter estimation: 
\begin{equation}
b = (X^TX)^{-1}(X^Ty)
\end{equation}

Allocation of variation: 
\begin{equation}
SSY=\sum_{i=1}^n y_i^2  
\end{equation}

\begin{equation}
SS0=n \overline y^2 
\end{equation}
\begin{equation}
SST=SSY-SS0
\end{equation}
\begin{equation}
SSE={y^T y-b^T X^Ty}
\end{equation}
\begin{equation}
SSR=SST-SSE
\end{equation}

Where \\ 
SSY=sum of squares of Y \\
\hspace{1cm} SST=total sum of squares \\
\hspace{1cm} SS0=sum of squares of y \\
\hspace{1cm} SSE=sum of squared errrors \\
\hspace{1cm} SSR= sum of squares given by regression \\

Coefficient of determination: 
\begin{equation}
R^2=\frac{SSR}{SST}=\frac{SST-SSE}{SST}
\end{equation}

Coefficient of multiple correlation 
\begin{equation}
R=\sqrt{\frac{SSR}{SST}}
\end{equation}

 The interaction term is ignored in this case, since the experiments described earlier in the paper have clearly indicated that there is no significant interaction between the predictor variables. Hence, intercept=0.
 
The ratio of Training data : Test data is 90-10 as the data available is less. However, the ratio could be changed to 80-20, 70-30 with the increasing size of the data available.

The elasticity values obtained exactly match with the values obtained earlier and holds good for the test data set as well. The R squared test conducted for validation yields $0.99$. This indicates excellent fit.

\begin{table}[!htpb]
\centering
 \begin{tabular}{|c|c|c|c|}
\hline
 & IRS & CRS & DRS\\
 \hline
SSY & 16984.7190 & 269.9263 & 217.7689\\
 \hline
SSO & 999.1011 & 269.5206 & 217.4285 \\
 \hline
 SST & 15985.6179 & .4056 & 0.3403\\
\hline
SSR & 15984.61683 & 0.4056 & 0.3374\\
\hline
R squared & .9999 & .9999 & .9913 \\
\hline
Alpha & 1.8 & 0.9 & 0.8 \\
\hline
Beta & 0.08 or 0.1 & 0.099 or 0.1 & 0.08 or 0.1\\
\hline
\end{tabular}
\caption{Multiple Linear Regression Results}
\end{table}

\section{Conclusion}
With the increase in utility computing , the focus has now shifted on cost effective data centers. Data centers are the backbone to any cloud environment that caters to demand for uninterrupted service with budgetary constraints. AWS and other data center providers are constantly improving the technology and define the cost of servers as the principle component in the revenue model. For example, AWS spends approximately 57\% of their budget towards servers and constantly improvise in the procurement pattern of three major types of servers. Here in this paper, we have shown how to achieve profit maximization and cost minimization within certain constraints. We have mathematically proved that cost minimization can be achieved at the phase of increasing return to scale, whereas profit maximization can be attained at the phase of decreasing return to scale. The Cobb Douglas model which is a special case of CES model  is used by the authors as revenue model which looks at such situation i.e include two different input variables for the costs of two different types of servers.  Thereafter Factor analysis done is \(2^2\), the four factors (S,I,P,N) were combined into two factors. This helped in overcoming the problem of curvature violation which is a disadvantage of the CD function. 

The factors, number of servers (S) and investment in infrastructure (I) were combined to ‘cost of deploying new server’. The other two factors cost of power (P) and networking cost (N) were combined to ‘cost of power and cooling’. Our work has established that the proposed model is in sync with and optimal output elasticities with real -time dataset. As server hardware is the biggest factor of total operating cost of data center and power is the most significant cost amongst other cost segments of data center, we have taken these two cost segments prominently in our output elasticity calculation.
The analytical exercise, coupled with a full factorial design of an experiment quantifies the contribution of each of the factors towards the revenue generated.  The take away factor for a commercial data centre from this paper is that the new server procurement and deployment cost plays a major role in the cost revenue dynamics. Also, that the response variable is a function of linear predictors.
The interaction factor has been ignored after performing factor analysis which subsequently showed the effect of interaction between factor 1 and factor 2 on revenue as less than 1\%, which is insignificant. Hence the two factors are independent from each other and have individual contribution towards the revenue. The authors have extensively calculated the contribution of each factor which is significant for the proposed model. All calculations have shown that new server cost is a major impact and hence with the technological progress and new techniques, the amount of investment on the deployment of new server could be reduced such that the profit can be maximized.\\

In the experiment including replications, the effects computed from a sample are random variables and would be different if another set of experiments is conducted. We have  calculated confidence interval for the effects for replication experiment, to check if a measured value is significantly different from zero. The results for CI , does not include zero in any intervals, so all effects are significantly different from zero at this confidence level. This implies that with 90\% cnfidence we can say that the actual values of effects are in the range as expected. 

 Since the prediction of the constant technological progress cannot be precisely modeled, the experiments performed taking the randomly generated data proves that the model used by the authors is valid to encompass the developments due to technological progress.
 \\
 
Our choice of elasticity depends on the dynamics between the factors and the benchmark constraints of optimization. CRS, for example requires the sum of the elasticities to equal 1 and DoE reveals that factor1 contributes to the response variable to a lesser extent compared to factor 2. Therefore, in order that revenue growth may be modeled in a balanced fashion, elasticity value for factor 1 has been set to much higher value compared to factor 2. The same phenomenon is observed in the cases of IRS and DRS and identical heuristic has been applied to pre-determine the choice of elasticities. The authors have verified the heuristic choices through fitting and regression. The results obtained from least square test (with k =1)or otherwise and multiple linear regression match closely with the assumptions made. Matlab codes where written using fmincon which attempts to find a constrained minimum of a scalar function of several variables starting at an initial seed of 0.4 and 0.1. The authors have performed subsequent calculations using active set solver to estimate the elastictity values which support the assumption made initially. The optimal revenue is obtained for all cases which are shown in the paper. So The authors conclude that by controlling the elasticity values the target revenue can be generated .
It may be checked easily that if one unit of cost is increased in infrastructure, the revenue goes up proportionately. There's a plausible explanation behind this. If infrastructure costs go up, which includes deployment of new servers, the revenue is supposed to go up; the data is taken from real world practicing data centers and all major players thrive on capacity building because business of cloud grows on the so called elasticity of service and unbounded time and space given to customers provided they pay differential fees. Therefore, investment ( cost ) positively affects the revenue! Indirectly, increase in cost implies more usage seems quite plausible. After all, these large data centres are not supposed to be functioning in a typical competitive market because of lumpiness (one time large non-fragmented outlay) in the fixed cost associated with establishing these centers. Small firms can't do it. So, the firms which end up doing this must get a premium for engaging with this business. It would be very interesting to use a Hirschman-Herfindahl Index (HHI) to find the level of concentration of firms in this business. Higher concentration means smaller number of firms controlling larger share of the business and vice versa. If HHI is high, meaning a few firms control the business, then new cost outlay raises revenue and that raises profit. If the profits are large and there is no entry barrier in this market, more firms enter and since business does not grow by the same rate necessarily (in the short run), the price falls and profit falls (at least does not rise). The authors would imagine that profits are surely positive but not too high (conversely, profits are high but entry into markets are restricted by technology or something else) so that it attracts too much attention and everybody starts investing in it. This is still the sunshine industry, so profits need to be higher regardless.  
The paper is potentially a good working tool for the entrepreneur empowering them with efficient/optimal resource allocation for all the inputs. There could be different combinations of resource allocation, even for a specific quantum of output. The concavity of the Production Possibility Curve ensures that. The proposed model has shown to be amenable to higher scale. Thus, any significant increase in the budget, consequently scale of investment on inputs, does no way invalidate the conclusion. Again, the fundamental law of variable proportions and its ultimate, the law of diminishing returns, has been found to be operative.

\ifCLASSOPTIONcaptionsoff
  \newpage
\fi

\vspace*{-2\baselineskip}
\begin{IEEEbiography}
    [{\includegraphics[width=1in,height=1.25in,clip,keepaspectratio]{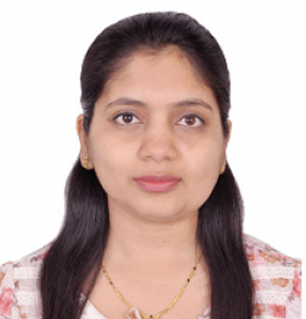}}]
{Ms. Gambhire Swati Sampatrao,} Faculty, PESIT-BSC holds Bachelors and Masters degree in Computer Science and Engineering. She has a teaching experience of 6 years in various technical institutions. Her research areas include Cloud Computing, Machine Learning and Optimization.
\end{IEEEbiography}
\vspace*{-4\baselineskip}
\begin{IEEEbiography}
[{\includegraphics[width=1in,height=1.25in,clip,keepaspectratio]{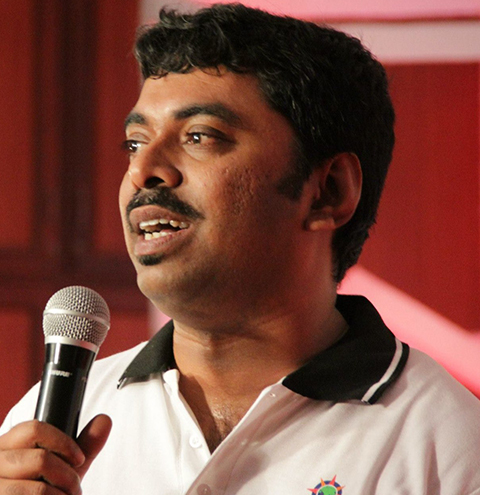}}]
{Dr. Snehanshu Saha,} Professor, PESIT-BSC was born in India, where he completed degrees in Mathematics and Computer Science. He went on to earn a Masters degree in Mathematical Sciences at Clemson University and Ph.D. from the Department of Mathematics at the University of Texas at Arlington
in 2008. After working briefly at his Alma matter, Snehanshu moved to the University of Texas El Paso as a regular full time faculty in the department of Mathematical Sciences, where he taught Differential equations, Advanced Calculus, Linear Algebra and Computational Mathematics. He is a Professor of Computer Science and Engineering at PESIT South Campus since 2011 and heads the Center for Applied Mathematical Modeling and Simulation. He has published 35 peer-reviewed articles in International journals and conferences and has been on the TPC of several IEEE R10 conferences. Has been IEEE Senior Member and ACM Professional Member since 2012.
   \end{IEEEbiography}
   \vspace*{-2\baselineskip}
   \begin{IEEEbiography}
      [{\includegraphics[width=1in,height=1.25in,clip,keepaspectratio]{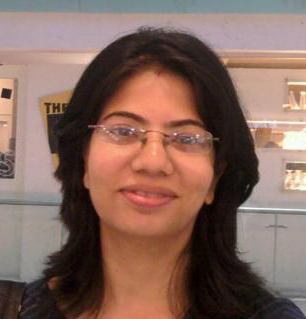}}]
    {Ms. Sudeepa Roy Dey,} Faculty, PESIT-BSC is a Bachelors and Masters degree holder in Computer Science and Engineering. She has eight years of teaching and research experience. Her research areas are Cloud Computing and Optimization techniques.
    \end{IEEEbiography}
     \vspace*{-12\baselineskip}
         \begin{IEEEbiography}
[{\includegraphics[width=1in,height=1.25in,clip,keepaspectratio]{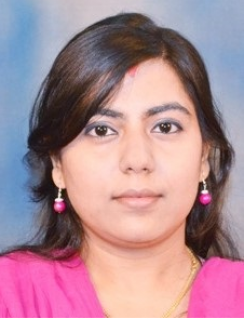}}]
    {Ms. Bidisha Goswami,*} Faculty, PESIT-BSC is working as an Assistant Professor of PES Institute of Technology Bangalore South Campus for last 4 yrs. She is perusing her Ph.D. in resource allocation in Service Computing. Her research interest includes Bio Inspired service computing,Intelligent Agent Technology, Cloud Computing and Optimization. 
    \end{IEEEbiography}
 \vspace*{2\baselineskip}
\begin{flushleft}
*Correspondence:
bidishagoswami@pes.edu,\\
Department of Computer Science and Engineering,\\ 
PESIT-BSC, Bangalore, India 560100,\\
Phone: 080-66186635.
\end{flushleft} 
\begin{IEEEbiography} [{\includegraphics[width=1in,height=1.25in,clip,keepaspectratio]{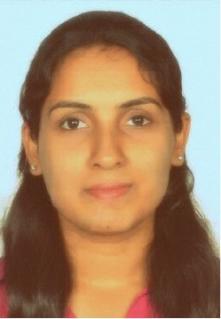}}]
   {Ms. Sai Prasanna M S,} Faculty, PESIT-BSC has B.E and M.Tech degrees in Computer Science and Engineering. She has teaching experience of two years. Her primary research interest is Cloud Computing and Optimization.
    \end{IEEEbiography}
     \vspace*{-12\baselineskip}
    
     \vspace*{-12\baselineskip}

\end{document}